\documentclass[runningheads]{llncs}
\usepackage[T1]{fontenc}
\usepackage{graphicx}
\usepackage{comment, booktabs, multirow, bbding, amssymb, amsmath}
\usepackage[normalem]{ulem}
\useunder{\uline}{\ul}{}
\begin{document}
\title{Multi-scale Scanning Network for Machine Anomalous Sound Detection}
%
%
\author{Yucong Zhang\inst{1,2}\orcidID{0009-0001-6553-3890} \and
Juan Liu\inst{1,3}$^\star$\orcidID{0000-0001-9344-7415} \and
Ming Li\inst{1,2}\thanks{Corresponding authors: Juan Liu and Ming Li}\orcidID{0000-0002-6406-1983}}
\authorrunning{Y. Zhang et al.}
%
\institute{School of Computer Science, Wuhan University, Wuhan, China \and
Suzhou Municipal Key Laboratory of  Multimodal Intelligent Systems, Digital Innovation Research Center, Duke Kunshan University, Suzhou, China \and School of Artificial Intelligence, Wuhan University, Wuhan, China \\
\email{\{yucong.zhang, ming.li369\}@dukekunshan.edu.cn}\\}
\maketitle              
\begin{abstract}
    Machine sounds exhibit consistent and repetitive patterns in both the frequency and time domains, which vary significantly across scales for different machine types. For instance, rotating machines often show periodic features in short time intervals, while reciprocating machines exhibit broader patterns spanning the time domain. While prior studies have leveraged these patterns to improve Anomalous Sound Detection~(ASD), the variation of patterns across scales remains insufficiently explored. To address this gap, we introduce a Multi-scale Scanning Network~(MSN) designed to capture patterns at multiple scales. MSN employs kernel boxes of varying sizes to scan audio spectrograms and integrates a lightweight convolutional network with shared weights for efficient and scalable feature representation. Experimental evaluations on the DCASE 2020 and DCASE 2023 Task 2 datasets demonstrate that MSN achieves state-of-the-art performance, highlighting its effectiveness in advancing ASD systems.
    
\end{abstract}
\section{Introduction}
Machine Anomalous Sound Detection~(ASD) aims to differentiate abnormal machine operating sounds from normal ones. Due to the scarcity of anomalies, ASD tasks often require models to detect abnormal samples without prior exposure to them~\cite{dcase2020,dcase2023}. In recent years, a variety of methods have been developed for the ASD task. Several generative methods have been found useful, aiming to model the distribution of the normal data by reconstructing audio spectrograms~\cite{Ellen2019Auto,Kaori2020IDNN}. However, their strong generalization capability can lead to the unintended reconstruction of anomalous samples~\cite{kuroyanagi2022improvement,zavrtanik2024anomalous}, resulting in detection failures. To overcome this limitation, Discriminative Representation Learning~(DRL) methods~\cite{giri2020mobilenetV2,zhang23j_interspeech,wilkinghoff2024self} have gained prominence. These approaches learn robust representations by classifying audio clips based on supplementary information like machine types, and have proven highly effective in recent DCASE challenges. However, they often require large amounts of annotated data, which can be difficult to collect or annotate.

To address the challenge of limited training samples in ASD, researchers have explored data augmentation and fine-tuning strategies to enhance DRL models. Data augmentation involves generating synthetic samples by introducing anomalies on audio spectrograms~\cite{zavrtanik2024anomalous,chen2023effective,tanaka2024few}, creating fake samples in latent spaces~\cite{wilkinghoff2024self,zeng2023geco}, or simulating machine sounds with diverse physical properties through finite element analysis~\cite{zhang2023data}. Pre-trained generative models like AudioLDM~\cite{liu2023audioldm} have also been used to generate machine sounds under varying conditions by translating operational attributes into textual descriptions~\cite{zhang2024first}. While these methods diversify training data, poorly designed synthetic samples risk degrading model performance.

Fine-tuning pre-trained models has emerged as a promising solution for few-shot ASD tasks. Recent studies demonstrate the effectiveness of contrastive learning in initializing DRL model weights using in-domain machine data, providing a robust starting point. A discriminative task is then employed to train the representation specifically for ASD~\cite{guan2023anomalous}. Notably, transferring weights pre-trained on large-scale speech data to ASD yields competitive results after fine-tuning on machine audio~\cite{han2024exploring}. To better align pre-trained models with the inductive bias of machine audio, researchers have fine-tuned models like BEATs~\cite{chen23beats} and CED~\cite{dinkel2024ced}, originally trained on large-scale datasets such as AudioSet~\cite{gemmeke2017audio}. This approach significantly enhances ASD task performance, achieving state-of-the-art (SOTA) results~\cite{jiang2024anopatch,zheng2024improving}. However, the fixed transformer-based architectures of these methods limit flexibility, posing challenges for adaptation and customization in ASD tasks. This limitation is particularly significant given the distinct spectrogram patterns identified in previous studies, which have proven promising for anomaly detection~\cite{LiuCQUPT2022,MaiDGB22}.

To automate the exploration of these spectrogram patterns, a range of methods has been proposed. For example, the multi-head self-attention mechanism~\cite{Vaswani2017AttentionIA} has been employed to adaptively filter log-Mel spectrograms~\cite{zhang23fa_interspeech}. Similarly, global weighted ranking pooling (GWRP)~\cite{kolesnikov2016seed} has been applied to the time domain of spectrograms~\cite{guan2023time}, adapting to different machine types. Additionally, squeeze-and-excitation modules and band-wise splitting strategy have been investigated
to capture both temporal and spectral patterns during model training~\cite{zhang2024fte}. The experimental results of all these approaches demonstrate the effectiveness of incorporating spectrogram pattern analysis into the ASD training processes.

\begin{figure*}[t]
	\centering
	\includegraphics[width=\textwidth]{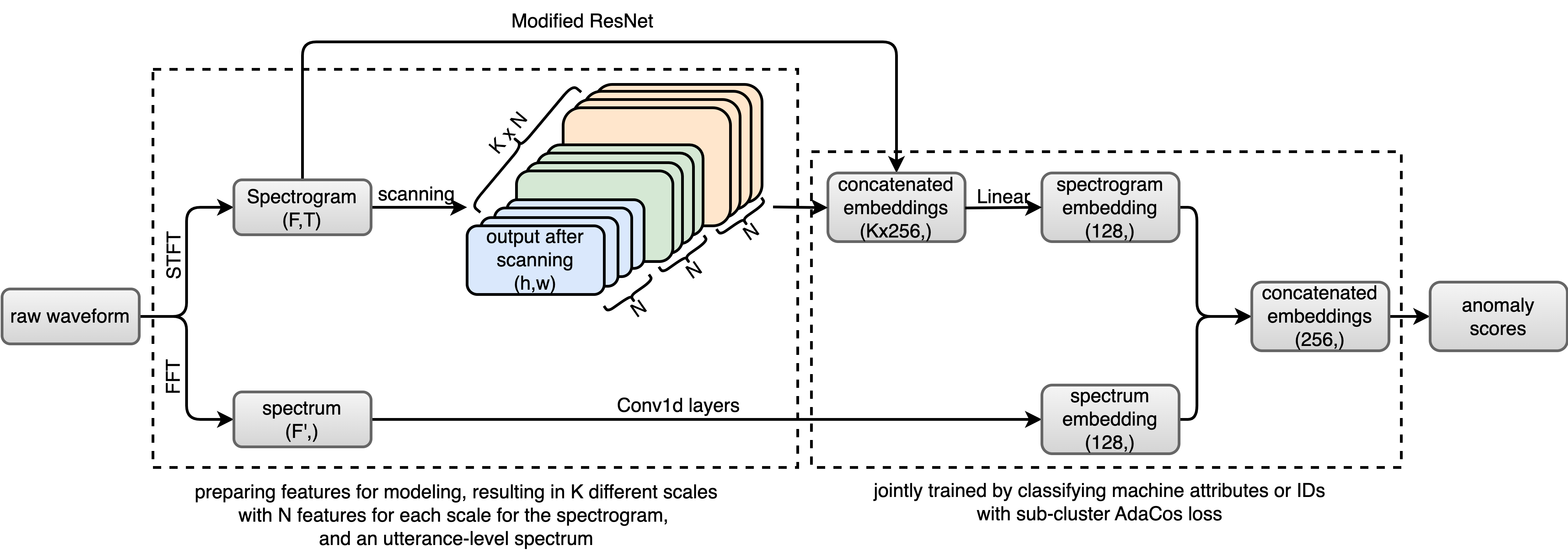}
	\caption{The overview of our proposed method. K is the number of kernel boxes used. N is the total number of features scanned by each kernel. F,T and F' are the dimensions of the spectrogram and spectrum computed from the input audio clip.}
	\label{fig:overall_structure}
\end{figure*}

While several studies have explored automatic feature extraction from machine spectrograms, few have addressed patterns across multiple scales. To address this gap, we propose a Multi-scale Scanning Network~(MSN) that utilizes multiple kernel boxes with different sizes to capture information across the entire spectrogram. The outputs from all kernel boxes are processed through a shared ResNet-based network~\cite{he2016deep}, and the resulting features are concatenated into a unified embedding, which is fed into auxiliary classification layers. By leveraging this multi-scale approach, our method effectively learns diverse patterns and enhances the representation of machine sound features. We evaluate our model on the DCASE 2020 and DCASE 2023 Task 2 benchmarks, demonstrating superior performance with the proposed module. Our implementation is publicly available on GitHub\footnote{Codes available at \url{https://github.com/yucongzh/MSN-Net}}.


\section{Proposed Method}

\subsection{Backbone}
Our method employs the widely-used dual-path structure as its backbone, as shown in Figure~\ref{fig:overall_structure}. Each path consists of a sub-network, and their outputs are concatenated to generate a unified embedding. The sub-network lying below in the figure processes the utterance-level spectrum, capturing magnitude information across the entire frequency range of the spectrogram. The second sub-network utilizes the magnitude spectrogram, preserving frequency information over time. This dual-path architecture has demonstrated strong performance in ASD tasks~\cite{wilkinghoff2024self,zhang2024fte,liu2022anomalous}. Its effectiveness may stem from the spectrum's ability to complement information potentially missing in the spectrogram, while high-frequency resolution proves essential for certain machine types.

\subsection{Spectrogram Encoding}
As depicted in the top section of Figure~\ref{fig:overall_structure}, the raw waveform is converted into an audio spectrogram using the Short Time Fourier Transform~(STFT). This transformation captures both time and frequency information, emphasizing local variations and characteristics within the audio signal. A modified ResNet~\cite{zhang2024fte} architecture is employed to process the spectrogram, as detailed in Table~\ref{tab:enhanced}. The modified ResNet integrates enhanced Squeeze-and-Excitation~(SE)~\cite{hu2018squeeze} modules, which assign dynamic weights not only across different channels but also along other dimensions. This design enables the model to effectively capture features across both the frequency and time axes, improving its representational capacity.

\subsection{Spectrum Encoding}
As illustrated in the lower section of Figure~\ref{fig:overall_structure}, the raw waveform is transformed into an utterance-level spectrum using Fast Fourier Transform~(FFT). This approach captures the overall frequency content of the audio signal, ensuring comprehensive representation of the spectral information. Following established methods for spectrum encoding~\cite{wilkinghoff2024self}, the spectrum undergoes processing through several 1D convolutional layers, followed by fully connected linear layers to derive the final feature representation. The detailed architecture of the spectrum encoding pathway is outlined in Table~\ref{tab:spectrum_pathway}.

\begin{table}[t]
\centering
\begin{minipage}[t]{0.48\textwidth}
\centering
\caption{Structure of the modified ResNet block shown in Figure~\ref{fig:overall_structure}. n indicates the number of layers or blocks, c is the number of output channel or dimension, k is the kernel size and s is the stride. This is used to encode the audio spectrogram.}
\label{tab:enhanced}
\resizebox{\columnwidth}{!}{
\begin{tabular}{@{}ccccc@{}}
\toprule
Operator     & n                  & c                  & k      & s      \\ \midrule
Modified SE  & -                  & -                  & -      & -      \\
Conv2d       & 1                  & 16                 & (7,7)  & (2,2)  \\
MaxPooling   & -                  & -                  & (3,3)  & (2,2)  \\
Modified SE  & -                  & -                  & -      & -      \\ \midrule
ResNet Block & \multirow{3}{*}{1} & 16                 & (3,3)  & (1,1)  \\
Modified SE  &                    & -                  & -      & -      \\
ResNet Block &                    & 16                 & (3,3)  & (1,1)  \\ \midrule
ResNet Block & \multirow{3}{*}{4} & (32, 64, 128, 256) & (3,3)  & (2,2)  \\
Modified SE  &                    & -                  & -      & -      \\
ResNet Block &                    & (32, 64, 128, 256) & (3,3)  & (1,1)  \\ \midrule
MaxPooling   & -                  & -                  & (h, w) & (h, w) \\ \bottomrule
\end{tabular}
}
\end{minipage}
\hfill
\begin{minipage}[t]{0.48\textwidth}
\centering
\caption{Structure of the Conv1d layers shown in Figure~\ref{fig:overall_structure} with the same notations shown in Table~\ref{tab:enhanced}. This is used to encode the utterance-level spectrum.}
\label{tab:spectrum_pathway}
\begin{tabular}{@{}ccccccc@{}}
\toprule
Operator     && n && c                   & k      & s      \\ \midrule
Conv1d && 1 && 128            & 256  & 64  \\
Conv1d && 1 && 128            & 64  & 32  \\
Conv1d && 1 && 128            & 32  & 4  \\
flatten  && - && -                   & -      & -      \\ \midrule
Linear   && 5 && 128                & -      & -      \\ \bottomrule
\end{tabular}
\vspace{3mm}
\centering
\caption{Structure of the convolution module for the multi-scale inputs with the same notations shown in Table~\ref{tab:enhanced}. This module convert the multi-scale features into embeddings with the same dimension.}
\label{tab:freq_pathway}
\begin{tabular}{@{}cccccc@{}}
\toprule
Operator     && n & c                   & k      & s      \\ \midrule
ResNet block && 2 & (32, 64)            & (3,3)  & (2,2)  \\
ResNet block && 1 & 64                  & (3,3)  & (1,1)  \\
StatsPool    && 1 & -                   & -      & -      \\ \midrule
Linear       && 1 & 1024                & -      & -      \\
Linear       && 1 & 256                 & -      & -      \\ \bottomrule
\end{tabular}
\end{minipage}
\end{table}

\subsection{Multi-scale Scanning Network}
In order to capture local features in the spectrogram of machine sounds, we design small multi-scale kernel boxes that scans the whole spectrogram along time and frequency domains, and put those scanned features together for modeling. The whole pipeline includes multi-scale scanning, and representation learning.

\subsubsection{Multi-scale Scanning}
We employ a collection of kernel boxes with varying dimensions, denoted as $\{K_{h \times w}\}$, where $h$ and $w$ represent the height and width of the kernel boxes, respectively. To simplify implementation, $h$ and $w$ are selected from powers of 2, ensuring computational efficiency and scalability across varying spectrogram resolutions. To emphasize frequency patterns, we use rectangular kernel boxes, with $h$ generally larger than $w$, to capture frequency-dominant features while maintaining time-domain granularity, motivated by the observation that machine sounds often exhibit distinct frequency characteristics Specifically, we define $K=12$ kernel boxes $\{K_{h \times w}\}$, with $h \in \{32, 64, 128, 256\}$ and $w \in \{16, 32, 64\}$. This diversity enables the model to focus on patterns of varying scales, effectively capturing both fine- and coarse-grained features within the spectrogram.

For each kernel box, $N_f$ scans are performed along the frequency axis and $N_t$ scans along the time axis, producing $N = N_f \times N_t$ features, each with fixed dimensions $(h, w)$. The scanning process is designed to ensure full coverage of the spectrogram by employing a calculated hop length, which determines the step size for sliding the kernel box. The hop length is calculated to balance coverage and computational efficiency, ensuring that no regions of the spectrogram are left unexamined while avoiding redundant computations. For kernel boxes of different scales, the step sizes for frequency (F\_step) and time (T\_step) are computed as follows.
Given an input spectrogram of dimensions $(F, T)$ and a kernel box of size $(h, w)$, the step sizes F\_step and T\_step are defined as:
$$
\text{F\_step} = 
\begin{cases} 
\max\left(1, \left\lfloor \frac{F - h}{N_f - 1} \right\rfloor\right) & \text{if } N_f > 1, \\
F - h & \text{otherwise.}
\end{cases}
$$

$$
\text{T\_step} = 
\begin{cases} 
\max\left(1, \left\lfloor \frac{T - w}{N_t - 1} \right\rfloor\right) & \text{if } N_t > 1, \\
T - w & \text{otherwise.}
\end{cases}
$$
The calculated step sizes dynamically adapt to the spectrogram dimensions and kernel box configurations, enabling the extraction of features that reflect the underlying spectral structure at each scale. This adaptability is crucial for handling spectrograms of varying resolutions and ensuring consistent feature extraction across different datasets. Each kernel box $K_{h \times w}$ extracts $N$ features from the input spectrogram. These features are stacked to form an $N$-channel feature map of dimensions $(N, h, w)$, as illustrated in Figure~\ref{fig:overall_structure}. The multi-scale structure allows the model to aggregate features from varying resolutions, enhancing its ability to capture localized and global patterns.

The output features from each kernel box are subsequently processed by a lightweight convolutional network. This network is specifically designed to extract scale-specific information while minimizing computational overhead, ensuring that the embedding captures essential spectral details without introducing significant latency. By integrating features from multiple scales, the model achieves a comprehensive representation of the input spectrogram, which is critical for downstream tasks such as classification or anomaly detection.

\subsubsection{Representation Learning}
A lightweight convolutional network integrates outputs from multi-scale scanning into a unified embedding, enabling efficient and scalable feature representation. The network, detailed in Table~\ref{tab:freq_pathway}, begins with an input of size~$(N, h, w)$, which is transformed into a lower-dimensional space through two residual blocks~\cite{he2016deep}. These blocks preserve critical spatial and contextual information while reducing dimensionality, leveraging skip connections to mitigate gradient vanishing and ensure stable training. A statistical pooling layer then computes channel-wise mean and standard deviation, producing a compact $(256,)$ embedding that captures robust, transformation-invariant statistics. If scanning is performed with $K$ multi-scale kernel boxes, $K$ embeddings are generated, each encapsulating information specific to a distinct scale. These embeddings are concatenated into a super embedding of size $(K \times 256,)$, which is refined by linear layers to produce the final spectrogram embedding, as visualized in Figure~\ref{fig:overall_structure}.

\subsection{Anomaly Detection}
Anomaly scores are calculated as the minimum cosine distance between prototypes of normal embeddings from the training dataset and test embeddings. For both DCASE 2020 and DCASE 2023 datasets, the same method is applied with different configurations. For DCASE 2020, scores are computed for each machine ID and type by comparing normal training samples with test samples of the same ID and type. For DCASE 2023, scores are computed for each machine type independently. Prototypes for each category are generated using K-Means clustering. For DCASE 2023, prototypes are created for both source and target domains, and the minimum cosine distance between the test sample and these prototypes is selected as the final anomaly score.

\begin{table}[htbp]
\caption{Model Comparison on the development test dataset of the DCASE 2020 Task 2 dataset. All results~(\%) are reported in terms of the mean of the AUC and pAUC. All the models are trained on only the normal machine sounds of the training dataset. "-" means that the result is not reported in the source paper.}
\label{tab:dc20_dev}
\resizebox{\textwidth}{!}{%
\begin{tabular}{@{}ccccccccc@{}}
\toprule
\multicolumn{9}{c}{DCASE 2020 Task 2} \\ \midrule
 &
  Pre- &
  \multicolumn{7}{c}{Development Dataset} \\
\multicolumn{1}{c|}{Models} &
  \multicolumn{1}{c|}{-trained} &
  Fan &
  Pump &
  Slider &
  T.Car &
  T.Conv &
  Valve &
  Mean \\ \midrule
\multicolumn{1}{c|}{2020 No.1~\cite{giri2020mobilenetV2}} &
  \multicolumn{1}{c|}{-} &
  80.65 &
  83.27 &
  93.41 &
  92.72 &
  {\ul \textbf{73.28}} &
  94.30 &
  86.27 \\
\multicolumn{1}{c|}{SC-AdaCos~\cite{wilkinghoff2021sub}} &
  \multicolumn{1}{c|}{-} &
  82.77 &
  81.81 &
  98.59 &
  {\ul \textbf{94.01}} &
  67.56 &
  96.63 &
  89.47 \\
\multicolumn{1}{c|}{MFN~\cite{hou2023decoupling}} &
  \multicolumn{1}{c|}{-} &
  83.71 &
  90.82 &
  98.70 &
  91.97 &
  71.29 &
  96.49 &
  88.83 \\
\multicolumn{1}{c|}{STgram~\cite{liu2022anomalous}} &
  \multicolumn{1}{c|}{-} &
  91.51 &
  86.85 &
  98.58 &
  91.06 &
  69.09 &
  99.04 &
  89.35 \\
\multicolumn{1}{c|}{ASD-AFPA~\cite{zhang23fa_interspeech}} &
  \multicolumn{1}{c|}{-} &
  95.51 &
  90.61 &
  99.04 &
  92.80 &
  70.35 &
  97.27 &
  90.93 \\
\multicolumn{1}{c|}{FTE-Net~\cite{zhang2024fte}} &
  \multicolumn{1}{c|}{-} &
  95.77 &
  94.99 &
  98.74 &
  93.98 &
  65.78 &
  99.62 &
  91.48 \\
\multicolumn{1}{c|}{Unsuper-TDGCN~\cite{yan2024transformer}} &
  \multicolumn{1}{c|}{-} &
  - &
  - &
  - &
  - &
  - &
  - &
  - \\ \midrule
\multicolumn{1}{c|}{CLP-SCF~\cite{guan2023anomalous}} &
  \multicolumn{1}{c|}{$\checkmark$} &
  95.11 &
  91.18 &
  98.65 &
  93.02 &
  69.00 &
  99.70 &
  91.12 \\
\multicolumn{1}{c|}{AnoPatch~\cite{jiang2024anopatch}} &
  \multicolumn{1}{c|}{$\checkmark$} &
  86.46 &
  93.10 &
  99.20 &
  96.10 &
  73.20 &
  97.53 &
  90.93 \\ \midrule
\multicolumn{1}{c|}{Ours} &
  \multicolumn{1}{c|}{-} &
  {\ul \textbf{98.98}} &
  {\ul \textbf{95.02}} &
  {\ul \textbf{99.59}} &
  90.99 &
  69.23 &
  {\ul \textbf{99.81}} &
  {\ul \textbf{92.27}} \\ \bottomrule
\end{tabular}
}
\end{table}

\begin{table}[htbp]
\caption{Model Comparison on the evaluation test dataset of the DCASE 2020 Task 2 dataset. All results~(\%) are reported in terms of the mean of the AUC and pAUC. All the models are trained on only the normal machine sounds of the training dataset. "-" means that the result is not reported in the source paper.}

\label{tab:dc20_eval}
\resizebox{\textwidth}{!}{%
\begin{tabular}{@{}ccccccccc@{}}
\toprule
\multicolumn{9}{c}{DCASE 2020 Task 2}                                                                                                                             \\ \midrule
                                                           & Pre-                              & \multicolumn{7}{c}{Evaluation Dataset}                           \\
\multicolumn{1}{c|}{Models}                                & \multicolumn{1}{c|}{-trained}     & Fan   & Pump  & Slider & T.Car          & T.Conv & Valve & Mean  \\ \midrule
\multicolumn{1}{c|}{2020 No.1~\cite{giri2020mobilenetV2}} &
  \multicolumn{1}{c|}{-} &
  89.42 &
  87.69 &
  93.68 &
  92.04 &
  \textbf{82.27} &
  93.51 &
  89.77 \\
\multicolumn{1}{c|}{SC-AdaCos~\cite{wilkinghoff2021sub}}   & \multicolumn{1}{c|}{-}            & 95.42 & 92.53 & 93.54  & 93.96          & 75.00  & 97.31 & 91.30 \\
\multicolumn{1}{c|}{MFN~\cite{hou2023decoupling}}          & \multicolumn{1}{c|}{-}            & 94.72 & 92.94 & 97.58  & 94.31          & 77.54  & 94.88 & 92.00 \\
\multicolumn{1}{c|}{STgram~\cite{liu2022anomalous}}        & \multicolumn{1}{c|}{-}            & -     & -     & -      & -              & -      & -     & -     \\
\multicolumn{1}{c|}{ASD-AFPA~\cite{zhang23fa_interspeech}} & \multicolumn{1}{c|}{-}            & -     & -     & -      & -              & -      & -     & -     \\
\multicolumn{1}{c|}{FTE-Net~\cite{zhang2024fte}}           & \multicolumn{1}{c|}{-}            & 99.72 & 94.78 & 98.17  & \textbf{94.61} & 69.50  & 93.52 & 91.72 \\
\multicolumn{1}{c|}{Unsuper-TDGCN~\cite{yan2024transformer}} &
  \multicolumn{1}{c|}{-} &
  88.08 &
  86.37 &
  98.11 &
  92.25 &
  79.68 &
  {\ul \textbf{99.87}} &
  90.73 \\ \midrule
\multicolumn{1}{c|}{CLP-SCF~\cite{guan2023anomalous}}      & \multicolumn{1}{c|}{$\checkmark$} & -     & -     & -      & -              & -      & -     & -     \\
\multicolumn{1}{c|}{AnoPatch~\cite{jiang2024anopatch}} &
  \multicolumn{1}{c|}{$\checkmark$} &
  95.56 &
  94.34 &
  {\ul 99.77} &
  {\ul 96.00} &
  {\ul 83.74} &
  96.26 &
  {\ul 94.28} \\ \midrule
\multicolumn{1}{c|}{Ours} &
  \multicolumn{1}{c|}{-} &
  {\ul \textbf{99.92}} &
  {\ul \textbf{96.61}} &
  \textbf{98.74} &
  94.13 &
  70.43 &
  93.41 &
  \textbf{92.21} \\ \bottomrule
\end{tabular}
}
\end{table}

\begin{table}[htbp]
\caption{Model Comparison on the development test dataset of the DCASE 2023 Task 2 dataset. All results~(\%) are reported in terms of the Harmonic Mean~(H.Mean) of the AUC and pAUC. All the models are trained on only the normal machine sounds of the training dataset. "-" means that the result is not reported in the source paper.}
\label{tab:dc23_dev}
\resizebox{\textwidth}{!}{%
\begin{tabular}{@{}cccccccccc@{}}
\toprule
\multicolumn{10}{c}{DCASE 2023 Task 2} \\ \midrule
 &
  Pre- &
  \multicolumn{8}{c}{Development Dataset} \\
\multicolumn{1}{c|}{Models} &
  \multicolumn{1}{c|}{-trained} &
  Bearing &
  Fan &
  G.Box &
  Slider &
  T.Car &
  T.Train &
  Valve &
  H.Mean \\ \midrule
\multicolumn{1}{c|}{2023 No.1~\cite{chen2023mdam}} &
  \multicolumn{1}{c|}{-} &
  64.41 &
  {\ul \textbf{76.27}} &
  74.78 &
  91.83 &
  51.66 &
  53.17 &
  65.18 &
  68.11 \\
\multicolumn{1}{c|}{FeatEx~\cite{wilkinghoff2024self}} &
  \multicolumn{1}{c|}{-} &
  - &
  - &
  - &
  - &
  - &
  - &
  - &
  66.95 \\
\multicolumn{1}{c|}{MS-D2AE~\cite{chen2024multi}} &
  \multicolumn{1}{c|}{-} &
  - &
  - &
  - &
  - &
  - &
  - &
  - &
  - \\
\multicolumn{1}{c|}{FTE-Net~\cite{zhang2024fte}} &
  \multicolumn{1}{c|}{-} &
  62.76 &
  73.01 &
  {\ul \textbf{75.97}} &
  88.00 &
  53.26 &
  53.56 &
  78.07 &
  67.04 \\ \midrule
\multicolumn{1}{c|}{Han et al.~\cite{han2024exploring}} &
  \multicolumn{1}{c|}{$\checkmark$} &
  57.10 &
  62.76 &
  67.52 &
  79.11 &
  {\ul 63.47} &
  57.35 &
  67.79 &
  64.31 \\
\multicolumn{1}{c|}{AnoPatch~\cite{jiang2024anopatch}} &
  \multicolumn{1}{c|}{$\checkmark$} &
  {\ul 70.43} &
  66.65 &
  58.67 &
  81.88 &
  58.78 &
  {\ul 67.16} &
  53.73 &
  64.24 \\
\multicolumn{1}{c|}{Zheng et al.~\cite{zheng2024improving}} &
  \multicolumn{1}{c|}{$\checkmark$} &
  - &
  - &
  - &
  - &
  - &
  - &
  - &
  65.11 \\ \midrule
\multicolumn{1}{c|}{Ours} &
  \multicolumn{1}{c|}{-} &
  \textbf{65.43} &
  67.96 &
  71.74 &
  {\ul \textbf{92.37}} &
  \textbf{55.06} &
  \textbf{58.86} &
  {\ul \textbf{83.02}} &
  {\ul \textbf{68.65}} \\ \bottomrule
\end{tabular}
}
\end{table}

\begin{table}[htbp]
\caption{Model Comparison on the evaluation test dataset of the DCASE 2023 Task 2 dataset. All results~(\%) are reported in terms of the Harmonic Mean~(H.Mean) of the AUC and pAUC. All the models are trained on only the normal machine sounds of the training dataset. "-" means that the result is not reported in the source paper.}
\label{tab:dc23_eval}
\resizebox{\textwidth}{!}{%
\begin{tabular}{@{}cccccccccc@{}}
\toprule
\multicolumn{10}{c}{DCASE 2023 Task 2} \\ \midrule
 &
  Pre- &
  \multicolumn{8}{c}{Evaluation Dataset} \\
\multicolumn{1}{c|}{Models} &
  \multicolumn{1}{c|}{-trained} &
  B.Saw &
  Grinder &
  Shaker &
  T.Dro &
  T.Nsc &
  T.Tan &
  Vacuum &
  H.Mean \\ \midrule
\multicolumn{1}{c|}{2023 No.1~\cite{chen2023mdam}} &
  \multicolumn{1}{c|}{-} &
  60.97 &
  65.18 &
  63.50 &
  55.71 &
  84.72 &
  60.72 &
  92.27 &
  66.97 \\
\multicolumn{1}{c|}{FeatEx~\cite{wilkinghoff2024self}} &
  \multicolumn{1}{c|}{-} &
  - &
  - &
  - &
  - &
  - &
  - &
  - &
  68.52 \\
\multicolumn{1}{c|}{MS-D2AE~\cite{chen2024multi}} &
  \multicolumn{1}{c|}{-} &
  - &
  - &
  - &
  - &
  - &
  - &
  - &
  66.54 \\
\multicolumn{1}{c|}{FTE-Net~\cite{zhang2024fte}} &
  \multicolumn{1}{c|}{-} &
  \textbf{59.81} &
  69.69 &
  {\ul \textbf{82.94}} &
  57.31 &
  \textbf{87.41} &
  \textbf{67.20} &
  88.31 &
  {\ul \textbf{71.27}} \\ \midrule
\multicolumn{1}{c|}{Han et al.~\cite{han2024exploring}} &
  \multicolumn{1}{c|}{$\checkmark$} &
  - &
  - &
  - &
  - &
  - &
  - &
  - &
  - \\
\multicolumn{1}{c|}{AnoPatch~\cite{jiang2024anopatch}} &
  \multicolumn{1}{c|}{$\checkmark$} &
  {\ul 69.71} &
  64.1 &
  80.3 &
  64.49 &
  85.04 &
  {\ul 72.6} &
  92.24 &
  74.23 \\
\multicolumn{1}{c|}{Zheng et al.~\cite{zheng2024improving}} &
  \multicolumn{1}{c|}{$\checkmark$} &
  67.67 &
  71.18 &
  82.87 &
  {\ul 71.73} &
  {\ul 95.97} &
  68.52 &
  {\ul 98.18} &
  {\ul 77.75} \\ \midrule
\multicolumn{1}{c|}{Ours} &
  \multicolumn{1}{c|}{-} &
  57.01 &
  {\ul \textbf{72.12}} &
  79.10 &
  \textbf{58.43} &
  86.16 &
  65.38 &
  \textbf{88.33} &
  70.43 \\ \bottomrule
\end{tabular}
}
\end{table}

\section{Experiments}
\subsection{Datasets}
The experiments were conducted using the Task 2 datasets from the DCASE 2020 and DCASE 2023 challenges~\cite{dcase2020,dcase2023}. Both datasets feature a development dataset and an evaluation dataset, each containing a training subset with only normal audio clips and a test subset with both normal and anomalous audio clips. These datasets are widely recognized in the ASD community, with DCASE 2020 focusing on standard ASD tasks and DCASE 2023 addressing ASD under domain shifts. The DCASE 2020 dataset includes six machine types, each with multiple machine IDs, while the DCASE 2023 dataset comprises 14 machine types with data from both source and target domains. The DCASE 2023 dataset also introduces domain shifts across machine types under varying operating conditions. Unlike DCASE 2021 and DCASE 2022, DCASE 2023 does not provide specific domain information; only machine attribute labels are disclosed. This setup better reflects real-world scenarios, where domains are not clearly defined, increasing the difficulty of the ASD task and emphasizing the need for DRL models to leverage additional attribute information effectively.

\subsection{Evaluation Metrics}
The evaluation metrics follow the official DCASE challenges~\cite{dcase2020,dcase2023}. Three commonly used metrics are adopted for evaluating the ASD performance in this paper: area under the receiver operating characteristic curve~(AUC), partial-AUC~(pAUC) and the integrated scores. AUC is divided into source AUC and target AUC for the data in separate domains for DCASE 2023 challenge. pAUC is calculated as the AUC over a low false-positive-rate (FPR) range [0, 0.1]. The integrated score is the mean~(for DCASE 2020) or harmonic mean~(for DCASE 2023) of AUC and pAUC scores across all machine types, which is the official score used for ranking.

\subsection{Implementation details}
We follow the data processing methodology outlined in~\cite{zhang2024fte,wilkinghoff2024self}. For the DCASE 2023 Task 2 dataset, audio clips are either repeated or truncated to a fixed duration of 18 seconds, the maximum length, to handle the variability in clip lengths across machine types. For the DCASE 2020 Task 2 dataset, the audio clips are kept in their original form, each lasting 10 seconds. All audio samples are sampled at 16 kHz. Spectrograms are generated using the Short-Time Fourier Transform (STFT) with a window size of 1024 and a hop length of 512. The utterance-level spectrum is derived by applying the Fourier Transform to the entire signal. The number of classes is based on the combined categories of machine types and machine IDs (DCASE 2020) or attributes (DCASE 2023). In our experiments, T\_step and F\_step are configured as 32 and 8, respectively.

For training, we employ the wave-level mixup strategy~\cite{zhang2018mixup}, with the mixup coefficient drawn from $\text{Beta}\sim(0.2,0.2)$. For non-mixup samples, label smoothing is applied with a coefficient sampled from $\text{Uniform}\sim(0,0.5)$. We use Sub-cluster Adacos~\cite{wilkinghoff2021sub} as the loss function. The model is optimized using the ADAM optimizer with a learning rate of 0.001, a batch size of 64, and trained for 100 epochs on a single NVIDIA GeForce RTX 3090 GPU.

\subsection{Baseline Systems}
We utilize previous SOTA models as baselines. For both DCASE 2020 and DCASE 2023, we adopt the top-performing systems from the challenges~\cite{giri2020mobilenetV2,chen2023mdam}, single models employing DRL training from scratch, and models with pre-trained weights. Widely recognized methods such as Sub-cluster Adacos (SC-Adacos)~\cite{wilkinghoff2021sub} and MobileFaceNet (MFN)~\cite{hou2023decoupling} are included as strong baselines. We also integrate models specializing in analyzing machine sound spectrogram patterns~\cite{wilkinghoff2024self,zhang23fa_interspeech,zhang2024fte,liu2022anomalous,yan2024transformer,chen2024multi}. Additionally, we include results from pre-trained models, such as those pre-trained on large-scale speech data~\cite{han2024exploring}, AudioSet~\cite{jiang2024anopatch,zheng2024improving}, and normal data from the DCASE 2023 training set~\cite{guan2023anomalous}. All models are trained or fine-tuned exclusively on normal machine sounds and evaluated on test datasets from both development and evaluation phases.


\subsection{Comparison between Non-pre-trained Models}
For the DCASE 2020 dataset, as shown in Table~\ref{tab:dc20_dev} and Table~\ref{tab:dc20_eval}, our method demonstrates significant improvements over existing approaches. On the development dataset, it achieves the highest scores for the fan, pump, and toy car categories, with a mean score of 92.27, surpassing all SOTA models, including those with pre-training. On the evaluation dataset, it excels in the fan, pump, and slider categories, achieving a mean score of 92.21 and outperforming all other advanced methods without pre-training. Our approach demonstrates consistent and robust performance across various machine types and IDs, validating its efficacy in the Anomalous Sound Detection (ASD) task without domain shifts. 

For the DCASE 2023 dataset, as presented in Table~\ref{tab:dc23_dev} and Table~\ref{tab:dc23_eval}, the performance of all methods declines significantly compared to their results on the DCASE 2020 dataset, likely due to domain shifts caused by varying operating conditions. Despite these challenges, our model outperforms all non-pre-trained methods on most machine types. On the development dataset, our model outperform all the other methods on all machine types except for fan and gearbox, achieving the highest harmonic mean score of 68.65. On the evaluation dataset, our model's performance is comparable to the SOTA model's, and surpass the performance from the top-ranked teams in the challenge by a large margin. To obtain an overall result, we calculate the harmonic mean across all the machine types across the test of both development and evaluation datasets. As a result, our model outperforms all the other non-pre-trained methods with a harmonic mean of 69.53. These results demonstrate the robustness of our model under unknown domain variations.

\subsection{Comparison with Pre-trained Models}
For the DCASE 2020 dataset, as shown in Table~\ref{tab:dc20_dev} and Table~\ref{tab:dc20_eval}, our method demonstrates significant improvements over existing approaches. On the development dataset, it achieves the highest scores for the fan, pump, and toy car categories, with a mean score of 92.27, surpassing all SOTA models, including those utilizing pre-training. On the evaluation dataset, our method excels in the fan, pump, and slider categories, achieving a mean score of 92.21 and outperforming all other advanced methods without pre-training. These results validate the efficacy of our approach in the Anomalous Sound Detection (ASD) task, showcasing its consistent and robust performance across various machine types and IDs, even in the absence of domain shifts.

For the DCASE 2023 dataset, as presented in Table~\ref{tab:dc23_dev} and Table~\ref{tab:dc23_eval}, the performance of all methods declines significantly compared to their results on the DCASE 2020 dataset, likely due to domain shifts caused by varying operating conditions. Despite these challenges, our model outperforms all non-pre-trained methods across most machine types. On the development dataset, it achieves the highest harmonic mean score of 68.65, outperforming all other methods on all machine types except for fan and gearbox. On the evaluation dataset, our model's performance is comparable to that of the SOTA pre-trained models and surpasses the top-ranked teams by a large margin. These results highlight the robustness of our model in handling unknown domain variations and further reinforce its effectiveness in the ASD task.

\subsection{Pre-trained vs. Non-pre-trained}
Table~\ref{tab:dc20_dev} to Table~\ref{tab:dc23_eval} show that adopting a pre-training and fine-tuning strategy can improve the ASD performance. However, compared to non-pre-trained models, the performance gains from pre-trained ones are not substantial. The results indicate that models designed to focus on spectrogram pattern analysis without pre-training can perform competitively well for the ASD task, even outperforming models pre-trained on speech data. This suggests that pre-training methods remain under-explored for ASD tasks. In the future, we plan to incorporate multi-scale spectrogram pattern analysis into model pre-training, which may lead to better pre-trained models for ASD tasks.

\begin{table}[t]
\centering
\caption{Results~(\%) on DCASE 2023 Task 2 Dataset using different scaling strategy. These values are the harmonic means of AUC and pAUC across all machine types.}
\label{tab:ablation}\small
\begin{tabular}{@{}cccc@{}}
\toprule
Scales Strategies    & Dev.   & Eval.  & All   \\ \midrule
no multi-scales      & 65.02 & 63.87 & 64.44 \\
fix T, F multi-scale & 67.29 & 65.33 & 66.30 \\
fix F, T multi-scale & 66.62 & 63.95 & 65.26 \\
Ours                 & \textbf{68.65} & \textbf{70.43} & \textbf{69.53} \\ \bottomrule
\end{tabular}
\end{table}

\subsection{Ablation Study}
Finally, we conduct ablation studies to discuss the impact of the multi-scale strategy by comparing it with a version that does not utilize the strategy. As shown in Table~\ref{tab:ablation}, incorporating the multi-scale scanning significantly improves the model's performance compared to the version without the multi-scale approach. This demonstrates the importance of capturing patterns across different scales for effective machine ASD. Additionally, when the scale is fixed in the frequency domain while varying scales only in the time domain, the performance is worse than when the scales are varied in the frequency domain and fixed in the time domain. This suggests that multi-scale variations in the frequency domain are more critical for learning meaningful patterns. Finally, applying multi-scale strategies to both the frequency and time domains yields the best results, which corresponds to our proposed method.

\section{Conclusion}
This paper introduces Multi-Scale Network~(MSN), a novel DRL approach to address the challenge of extracting multi-scale spectrogram patterns for ASD. By employing kernel boxes of varying sizes and leveraging a lightweight convolutional network with shared weights, the MSN effectively captures unique characteristics of machine sounds across different scales. Experimental results on the DCASE 2020 and DCASE 2023 Task 2 datasets demonstrated that the proposed method achieves SOTA performance, highlighting its effectiveness, and potential for the ASD task. 

\section{Acknowledgements}
This research is funded in part by the Science and Technology Program of Suzhou City (SYC2022051). Many thanks for the computational resource provided by the Advanced Computing East China Sub-Center.

%
%
%
\bibliographystyle{splncs04}
\bibliography{mybib}
%




\end{document}